\documentclass[a4paper,UKenglish,cleveref,autoref,thm-restate]{lipics-v2021} 
\pdfoutput=1

\usepackage{cite}
\usepackage{amsfonts}
\usepackage{algorithmic}
\usepackage{makecell}
\usepackage{url}
\usepackage{xurl}
\usepackage{tikz}
\usetikzlibrary{arrows.meta, positioning}

\title{Modeling Dependency-Propagated Ecosystem Impact of Changes in Maintenance Activities: Evaluating Support Strategies in the PyPI Network}

\titlerunning{Modeling Dependency-Propagated Ecosystem Impact in PyPI}

\author{Alexandros Tsakpinis}
{fortiss GmbH, Munich, Germany}
{alexandros.tsakpinis@tum.de}
{https://orcid.org/0000-0001-6561-2866}
{}

\author{Emil Schwenger}
{fortiss GmbH, Munich, Germany}
{emil.schwenger@tum.de}
{https://orcid.org/0009-0009-6557-5095}
{}

\author{Alexander Pretschner}
{Technical University of Munich, Munich, Germany}
{alexander.pretschner@tum.de}
{https://orcid.org/0000-0002-5573-1201}
{}

\authorrunning{A. Tsakpinis, E. Schwenger, and A. Pretschner}

\Copyright{Alexandros Tsakpinis, Emil Schwenger, and Alexander Pretschner}



\relatedversion{}

\acknowledgements{}
\category{Technical Track Paper}

\hideLIPIcs 

\EventEditors{Robert Feldt, Maria Paasivaara, Daniel Mendez, Stefan Wagner, and Marvin Mu\~{n}oz Bar\'{o}n}
\EventNoEds{5}
\EventLongTitle{20th International Symposium on Empirical Software Engineering and Measurement (ESEM 2026)}
\EventShortTitle{ESEM 2026}
\EventAcronym{ESEM}
\EventYear{2026}
\EventDate{October 8--9, 2026}
\EventLocation{Munich, Germany}
\EventLogo{}
\SeriesVolume{394}
\ArticleNo{2}

\begin{document}
\nolinenumbers
\maketitle

\begin{abstract}
    \textbf{Background:}~Open source software ecosystems exhibit dense dependency networks in which maintenance degradation of structurally central packages can propagate widely. Despite increasing attention to open source sustainability, existing support mechanisms lack an explicit, dependency-aware notion of ecosystem-level impact to guide support decisions.
    \textbf{Aims:}~In this paper, we introduce a dependency-aware model of ecosystem impact that captures how changes in maintenance activities propagate through the Python Package Index (PyPI) ecosystem and affect its overall state. Based on this model, we prioritize packages for ecosystem support using our dependency-propagated notion of ecosystem impact.
    \textbf{Method:}~Applying this framework to a snapshot of 718,750 PyPI packages and over 2 million dependencies, we compare our impact-driven support strategy with existing support mechanisms (Tidelift, Ecosyste.ms, and GitHub Sponsors) and with PageRank as a baseline measure of structural importance.
    \textbf{Results:}~Our results show that a large share of the modeled ecosystem impact (approximately 80\%) can be attributed to just 0.1\% of all PyPI packages when prioritized based on dependency-propagated impact. In contrast, externally defined support sets vary substantially in their alignment with ecosystem impact. We further analyze maintainer reach and metadata accessibility, revealing that ecosystem impact, social footprint, and operational feasibility represent distinct but complementary dimensions of ecosystem support.
    \textbf{Conclusions:}~Dependency-aware ecosystem impact modeling provides a transparent and systematic basis for prioritizing support in large-scale software ecosystems. Our findings suggest that effective support strategies, driven by ecosystem stewards, funding bodies, and organizations operating support programs, should complement existing allocation logic with impact-informed decision making.
\end{abstract}
\section{Introduction}
\label{sec:introduction}

Over the past decades, Open Source Software (OSS) has become a fundamental component of modern software systems, with estimates suggesting that 80\% to 90\% of commercial software relies on open source components~\cite{ebert2008open, pittenger2016open, oss2022}. Within this landscape, reusable software libraries play a central role. Typically distributed as packages through ecosystem registries, they provide specialized functionality that reduces development effort~\cite{bauer2012structured}. Integrated as dependencies, they collectively form large-scale, layered dependency networks~\cite{cox2019surviving}. This structural interdependence introduces systemic risks: vulnerabilities or failures in a single package can propagate through direct and transitive dependencies, affecting large parts of the ecosystem~\cite{decan2018impact}, as illustrated by incidents such as Log4j~\cite{log4j2021}, LiteLLM~\cite{litellm}, and XZ Utils~\cite{xz2024}. While OSS communities often respond quickly to security issues~\cite{rahkema2022swiftdependencychecker}, sustained maintenance is not guaranteed. Libraries may experience declining activity or be abandoned entirely~\cite{bauer2012structured, raemaekers2011exploring, miller2023we}. In such cases, insufficiently maintained dependencies can propagate risks across the ecosystem, impacting potentially thousands of downstream projects~\cite{kula2014visualizing, tsakpinis2023analyzing}.

However, while maintenance-related risks in software libraries are well recognized~\cite{raemaekers2011exploring, miller2023we} and the propagation of risks through dependency networks is well established~\cite{kikas2017structure, fritz2024vulnerability}, there is no formalization that combines these perspectives into a dependency-aware notion of ecosystem-level impact, which quantifies how improving or degrading the maintenance activities of a given library affects the overall state of the ecosystem. This limitation is particularly relevant in the context of OSS support mechanisms, which are essential for sustaining maintenance through financial and non-financial interventions~\cite{tidelift2024, medappa2023sponsorship, linaaker2024sustaining}. A key challenge for these support mechanisms is the allocation of limited resources across libraries. However, existing support mechanisms operate without a notion of ecosystem impact~\cite{shimada2022sponsors, overney2020donations, zhou2022opencollective, conti2023incentivizing}. As a result, it remains unclear which subset of libraries should be prioritized to maximize ecosystem impact~\cite{pfeiffer2021critical, decan2019empirical, valiev2018ecosystem}. Beyond this structural perspective, additional challenges arise when operationalizing ecosystem support. The maintainer populations behind structurally critical libraries remain insufficiently characterized, limiting our understanding of how support strategies translate into social reach~\cite{casari2021open, zimmermann2019small, pfeiffer2021critical, avelino2016novel, coelho2017fail}. Furthermore, even when relevant libraries can be identified, practical constraints such as missing metadata (e.g.,~contact information, donation links, or repository URLs) can hinder maintainer outreach and support allocation~\cite{bommarito2019empirical, tsakpinis2026analyzing, tsakpinis2025analyzing, tsakpinis2024analyzing}. Finally, existing support mechanisms such as Tidelift, Ecosyste.ms, and GitHub Sponsors have not been systematically evaluated with respect to their dependency-propagated ecosystem impact, with available reports typically focusing on project-level improvements rather than transitive ecosystem-wide effects~\cite{tidelift_blog_paying_maintainers}. To guide our empirical analysis, we formulate the following research questions (RQs):

\textbf{RQ1:} When support is concentrated on libraries with the highest ecosystem impact on PyPI, how many libraries are required and what proportion of the ecosystem is affected?

\textbf{RQ2:} Given the libraries selected in RQ1, how does the selection translate into the number of maintainers reached?

\textbf{RQ3:} Given the libraries selected in RQ1, to what extent are maintainers reachable and supportable through PyPI and repository metadata, and how frequently are packages excluded from the analysis due to missing repository links?

\textbf{RQ4:} How do libraries associated with existing support mechanisms compare to the selection identified in RQ1 under the same analytical dimensions used in RQ1--RQ3?

We address these research questions by introducing a dependency-aware definition of ecosystem impact that models how changes in maintenance activities propagate through package dependency networks. We ground our analysis in the Python Package Index (PyPI), a widely used ecosystem in modern software development characterized by strong library integration patterns~\cite{decan2016topology, abdalkareem2020impact, Octoverse2025}. At a high level, ecosystem impact captures how changes in the maintenance state of a single package influence the overall state of the ecosystem. We operationalize this by modeling the ecosystem as a set of interdependent packages, where maintenance changes at the package level propagate through the dependency network and contribute to a global ecosystem state. By quantifying how individual packages affect this global state, we obtain a measure of their relative impact on the ecosystem. Aggregating these effects yields a global view that enables comparison across packages. Building on this definition, we frame ecosystem support as a constrained selection problem: given a target share of overall ecosystem impact, which packages should be prioritized for support? Using this framework, we analyze an impact-driven support allocation strategy that prioritizes libraries based on their dependency-propagated ecosystem impact. We examine how this strategy translates into the number and concentration of supported packages, the number of maintainers reached, and the availability of metadata required for delivering financial and non-financial support. Finally, we benchmark existing support mechanisms (Tidelift, Ecosyste.ms, and GitHub Sponsors) against our impact-driven selection strategy.

In response to the identified gaps, this paper makes the following contributions: First, we introduce a dependency-aware definition of ecosystem impact that models how simulated maintenance changes propagate through package dependency networks. Second, we operationalize ecosystem support allocation as a constrained selection problem and analyze an impact-driven prioritization strategy. Third, we extend the evaluation of support strategies beyond ecosystem impact by integrating maintainer reach and metadata accessibility as complementary dimensions of practical feasibility. Fourth, we provide a benchmark of existing support mechanisms (Tidelift, Ecosyste.ms, and GitHub Sponsors) against our dependency-propagated impact model, enabling comparison between impact-driven and externally defined allocation logics. Fifth, we systematically compare our proposed impact metric to PageRank as a baseline measure of structural importance, identifying when structural centrality serves as a suitable proxy for ecosystem impact and where it fails to capture the effects of maintenance changes and their propagation. Our contributions provide a dependency-aware basis for evaluating and prioritizing ecosystem support and may inform decision-making of ecosystem stewards, funding bodies, and organizations operating support programs.
\section{Background and Related Work}

\subsection{Open Source Sustainability and Support Mechanisms}

Open source software constitutes critical infrastructure in modern software supply chains, yet its sustainability is constrained by limited maintainer capacity and the often unpaid nature of maintenance work. Recent work conceptualizes sustainability as maintaining the socio-technical conditions that allow projects to remain healthy over time, emphasizing the need to support the ``human infrastructure'' behind OSS (e.g., reducing burnout and enabling stable maintenance routines)~\cite{linaaker2024sustaining}. In dependency-rich package ecosystems, sustainability also has a systemic dimension: the maintenance status of upstream libraries can affect many downstream projects through direct and transitive dependencies~\cite{valiev2018ecosystem}. To address these risks, a range of support mechanisms has emerged. Institutional support includes foundations that provide legal, organizational, and operational assistance, although prior work shows that such institutionalization complements rather than replaces project-level engineering and governance practices~\cite{izquierdo2018foundations}. Financial mechanisms include donations, crowdfunding, and sponsorship platforms. Empirical studies show heterogeneous outcomes: donations are used for diverse purposes but do not consistently translate into increased development activity~\cite{overney2020donations}, while evidence from Open Collective highlights diverse donation and spending patterns across projects~\cite{zhou2022opencollective}. Similarly, studies of donation platforms suggest that their effectiveness depends strongly on project characteristics such as size, maturity, and community engagement~\cite{zhang2025exploring}. Platform-mediated sponsorship mechanisms such as GitHub Sponsors provide another funding channel and shape developer activity and social dynamics in OSS ecosystems~\cite{shimada2022sponsors}.

Beyond these academic perspectives, industry and public-sector initiatives increasingly use dependency information to guide funding decisions. For example, Tidelift distributes subscription revenue based on their customer's dependency usage~\cite{tidelift_lifting_2023}. Ecosyste.ms has introduced ecosystem-level funds that allocate sponsorship to widely reused projects~\cite{ecosystms_funds_2025}, while public-sector programs such as the Sovereign Tech Agency fund open digital infrastructure to strengthen resilience and security~\cite{sta_2025}. Additional initiatives aim to normalize corporate contributions (e.g., Open Source Pledge)~\cite{opensourcepledge_2026} or support dependency-aware giving (e.g., thanks.dev)~\cite{thanksdev_canonical_2025}.
Recent work has also begun to systematize the evaluation of public funding interventions in open source ecosystems by proposing toolkits for measuring their impact~\cite{osborne2024toolkit}.

Despite the diversity of support mechanisms, none considers in their library selection how dependency networks amplify downstream effects of insufficient maintenance, limiting the assessment of ecosystem-level impact. This motivates our ecosystem-aware selection strategy that combines dependency-propagated impact and maintenance signals.

\subsection{Maintenance Activities and Project Health}

OSS project health is frequently operationalized through maintenance activities and the risk of inactivity, abandonment, or failure. Broader OSS health assessments combine multiple indicators, many of which are represented by metrics developed by the \textbf{C}ommunity \textbf{H}ealth \textbf{A}nalytics in \textbf{O}pen \textbf{S}ource \textbf{S}oftware (CHAOSS) project~\cite{linaaker2026assessing}. Empirical studies indicate that project failure or deprecation is multi-causal and influenced by socio-technical practices. For example, prior work analyzed deprecated GitHub projects and identified multiple factors contributing to project abandonment, including limited maintainer availability, project obsolescence, and competition from alternative solutions~\cite{coelho2017fail}. More broadly, recent work emphasizes that maintenance capacity is constrained by maintainer resources and the socio-technical conditions supporting maintainer labor~\cite{linaaker2024sustaining}. In package ecosystems, maintenance must also be considered in relation to dependency structure. A study of the PyPI ecosystem found that dependency-network properties and organizational support significantly influence sustained project activity~\cite{valiev2018ecosystem}. This implies that upstream packages with many dependents carry ecosystem-wide maintenance obligations. Given the role of dependency structure in maintenance, prior work investigated automated monitoring of maintenance activities for direct and transitive dependencies, highlighting limitations of approaches that ignore transitive relationships~\cite{tsakpinis2023analyzing}. Closely related to this limitation, existing work on maintenance activities of libraries has not systematically investigated how maintenance changes propagate through dependency networks and affect the ecosystem as a whole, motivating our formalization of a dependency-propagated notion of ecosystem impact that incorporates maintenance signals.

\subsection{Maintainer Reach and Metadata Accessibility}

Beyond identifying impactful libraries, the practical feasibility of ecosystem support depends on the availability of actionable maintainer and funding information. Missing or outdated metadata (e.g., repository URLs, donation links, or contact channels) can limit empirical analysis and hinder interventions such as sponsorship or coordinated support with maintainers~\cite{bommarito2019empirical}. Several prior studies examine these constraints in the PyPI ecosystem. Analyses of repository-link accessibility show that a non-trivial share of packages lacks valid repository links~\cite{tsakpinis2024analyzing}. Studies of donation-platform usage report limited adoption of donation links across projects~\cite{tsakpinis2025analyzing}. Both findings indicate that metadata availability tends to increase with structural importance and for packages appearing in dependency chains. Additional work on maintainer reachability shows that e-mail contact information is broadly available across PyPI, with only a small share of invalid or missing entries~\cite{tsakpinis2026analyzing}. To explain why such metadata gaps persist for repository and donation-platform links, prior work reports survey-based evidence on barriers that maintainers face when providing such metadata~\cite{tsakpinis2026authors}.~Together, these findings motivate metadata accessibility as an important feasibility dimension of ecosystem support, which we evaluate across the package sets produced by different support strategies.

\subsection{Structural Importance in Dependency Networks}

Package ecosystems can be modeled as directed dependency networks in which nodes represent packages and edges represent dependency relations. In such networks, packages may influence large portions of the ecosystem through transitive dependencies. Empirical studies show that these dependency graphs evolve over time and exhibit strong concentration effects, where relatively few packages accumulate many dependents~\cite{decan2019empirical}. Consequently, importance in package ecosystems should be assessed at the ecosystem level rather than relying solely on project-local popularity metrics. Structural importance is also closely related to systemic risk. Prior work on vulnerabilities in dependency networks demonstrates that issues in upstream packages can propagate widely to downstream projects, highlighting the need for ecosystem-level reasoning when analyzing propagation and mitigation~\cite{decan2018impact}. Similar dynamics apply beyond security, as maintenance degradation or abandonment in upstream dependencies can impose huge downstream costs~\cite{valiev2018ecosystem}. Quantifying such propagated effects requires measuring the transitive influence of packages within the dependency network. Centrality measures provide a natural baseline for this purpose, with PageRank capturing global influence based on the network’s link structure and serving as a proxy for transitive importance~\cite{brin1998anatomy}. Prior studies have applied PageRank to identify critical projects in package ecosystems, often in combination with maintainer-risk indicators such as the Truck Factor~\cite{pfeiffer2021critical}. Additionally, research suggests that trends in structural centrality can help identify declining packages earlier than traditional popularity metrics and support the identification of alternatives~\cite{mujahid2021toward,mujahid2023go}. 
Beyond academic studies, initiatives such as the OpenSSF Criticality Score identify critical open source projects by aggregating repository-level signals including activity, community engagement, and usage proxies~\cite{openssf_criticality_score}. While such approaches provide a practical notion of project criticality, their aggregation limits the ability to focus on specific dimensions such as maintenance dynamics. 
Taken together, while established importance measures such as PageRank are capable of capturing the potential influence of packages within a dependency network and prioritizing highly influential libraries, they lack the ability to capture how changes in maintenance state affect the ecosystem. Building on this observation, we use PageRank as a baseline measure of structural importance in the PyPI dependency network and compare it with our proposed impact metric for prioritizing libraries for support. This comparison allows us to analyze when PageRank serves as a suitable proxy for ecosystem impact and where it fails to capture the effects of maintenance changes and their propagation.
\section{Methodology}
\label{sec:methodology}

We conduct a large-scale, quantitative, observational study of the PyPI ecosystem based on mining, integrating, and analyzing package- and repository-level metadata. Our methodology combines dependency network analysis, repository-derived maintenance indicators, and ecosystem-level importance measures. We evaluate alternative ecosystem support strategies in terms of the number and concentration of supported packages, maintainer reach, supportability via contact and donation channels, and metadata-driven exclusion effects.

\subsection{Data Collection and Preprocessing}
\label{sec:data-collection}

To study ecosystem support strategies in the PyPI ecosystem, we collected a snapshot of all packages together with their dependency information and associated package- and repository-level metadata. Data collection was implemented as a reproducible pipeline, which combines multiple data sources and integrates them at package level into a consolidated~dataset.

\subsubsection{PyPI Package Discovery and Filtering}

We collected a comprehensive list of Python packages using an official API endpoint~\cite{pypi-simple}. Each package listed at the time of data collection was treated as a single unit of analysis. Packages that could not be resolved through the detailed PyPI endpoint~\cite{pypi_json_api} or that appeared only as dependency references without a package entry in the initial package list were excluded from the dataset. This preprocessing step ensures that all analyzed packages are associated with consistent dependency information. The resulting package list defines the scope of the ecosystem and serves as the baseline for all analyses and selection strategies.

\subsubsection{PyPI Package Metadata}

For each package, we retrieved package-level metadata using an official PyPI API~\cite{pypi_json_api}. These metadata include project URLs, which are used to identify source code repositories and donation platform links, as well as contact information of authors and maintainers in the form of available e-mail addresses. To protect the privacy of individuals, whenever valid contact information is available, we store only a reference to the corresponding project page rather than the e-mail address itself. In addition, we extracted direct and optional dependencies for the latest available version of each package. This dependency information is used to construct the PyPI dependency network, modeled as a directed graph with unrestricted transitive depth, which enables the computation of package importance based on PageRank. 
As described in more detail in Section~\ref{sec:impact-operationalization}, PageRank serves as a baseline for structural importance, against which we compare our dependency-propagated ecosystem impact metric. The retrieved PyPI metadata, including contact information, donation platform links, and repository URLs, is used for the analysis of maintainer reachability, supportability, and systematic exclusion effects in RQ3.

\textbf{Repository-level metadata:}
For packages with resolvable GitHub repositories, we collected repository-level metadata using the GitHub API. This includes repository ownership data (individual user vs.\ organization), and, where available, contact information listed at the repository owner level. To protect an individual's privacy, we do not store raw e-mail addresses but retain only references to the corresponding repository owner pages. In addition, we collected linked donation platform URLs, including GitHub Sponsors and externally referenced funding links. For organization-owned repositories, we collected associated members to quantify the number of maintainers reached by different package selection strategies, central to RQ2. The presence of contact information and donation links enables assessing maintainer reachability and supportability via repository metadata, relevant to RQ3.

\subsubsection{Maintenance Indicator}

To assess maintenance-related characteristics of packages, we used the \emph{Maintained} check from the OpenSSF Scorecard~\cite{openssf_scorecard}, a framework widely adopted across software repositories in academia and industry~\cite{scorecard-action-dependents, prominent-scorecard-users, tsakpinis2026forecasting}. This check provides a score from $0$ to $10$ based on commit and issue activity during the past 90 days. We selected this bounded and consistently computable signal for controlled large-scale comparisons of recent maintenance activity, not as a comprehensive measure of project health.

\subsubsection{Comparison of Impact Metric with Existing Importance Measures}
\label{sec:impact-operationalization}

To systematically compare our notion of dependency-propagated ecosystem impact with established ecosystem-level importance measures in the context of prioritizing libraries for support, we consider two complementary importance measures:
(i) a structural importance measure based on PageRank derived from the dependency network, and 
(ii) a usage-based importance measure based on package download counts obtained from the pepy.tech API~\cite{pepy-tech}. While both measures are retained in our dataset, PageRank is used as a baseline measure of structural importance against which we compare our proposed impact metric. This choice is motivated by conceptual, institutional, and practical considerations.

First, usage-based popularity and structural centrality capture different aspects of ecosystem importance. In our dataset, the two measures exhibit only a low-to-moderate association, with a Spearman rank correlation of $r_s = 0.31$, indicating that they frequently differ in their ranking and should not be treated as interchangeable signals~\cite{de2016comparing}. Structural centrality reflects the extent to which packages are relied upon by others and provides a natural baseline for assessing dependency-mediated influence in the ecosystem~\cite{brin1998anatomy}.
Second, PyPI deprecated download statistics from its official infrastructure due to concerns regarding data reliability and interpretation, including the influence of automated downloads and caching effects~\cite{pypi_downloads_deprecated}. In alignment with this decision, we avoid using download counts as a primary basis for ecosystem importance and instead use PageRank as a baseline for our comparison.
Third, PageRank provides practical advantages for large-scale ecosystem studies, as it can be computed solely from dependency information without requiring costly third-party~services~\cite{pepy-tech}.

\subsection{Data Analysis: Modeling dependency-propagated Ecosystem Impact}
\label{sec:data-analysis}

All analyses are part of a reproducible workflow implemented as a dashboard web application. Conceptually, the analysis evaluates an ecosystem support strategy based on dependency-propagated ecosystem impact by selecting sets of packages and examining the resulting implications in terms of package set size and concentration~(RQ1), maintainer reach~(RQ2), metadata accessibility~(RQ3), and alignment with existing support mechanisms~(RQ4).

As motivated in Section~\ref{sec:introduction}, maintenance changes in individual packages can propagate through dependency networks and affect large parts of the ecosystem. To reason about such system-wide effects, we require a measure that captures how local maintenance changes translate into ecosystem-level consequences. We therefore define \emph{ecosystem impact} as the dependency-propagated effect of changes in a package’s maintenance state. To enable comparison across packages, this effect is normalized with respect to the total dependency-propagated impact in the ecosystem. This definition reflects two key requirements: first, maintenance changes should propagate through direct and transitive dependencies, and second, their effect should scale with the extent to which a package influences other parts of the ecosystem. We use a linear additive baseline that treats transitive dependency occurrences equally to keep attribution transparent and scalable without imposing unvalidated weights. Richer models could distinguish direct and indirect dependencies to account for dependency depth, trace propagation through method-level call graphs, and represent structural relations such as imports, inheritance, and function calls~\cite{shen2025understanding,tang2025synfix}.

\begin{figure}[ht]
    \centering
    \includegraphics[width=\textwidth, trim=0mm 0mm 0mm 0mm, clip]{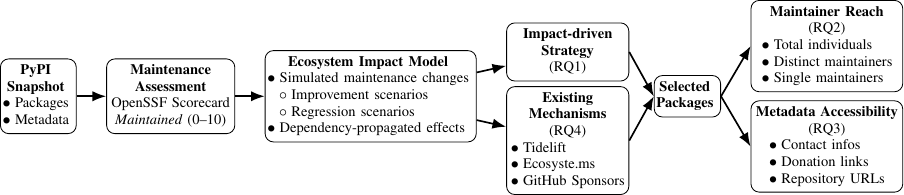}
    \caption{Conceptual overview of the impact-driven ecosystem support analysis framework}
    \label{fig:framework_overview}
\end{figure}

Concretely, we model the ecosystem as a set $\mathcal{P}$ of interdependent packages and define the overall ecosystem maintenance state $M_{\text{eco}}$ as the sum of maintenance levels of all packages, i.e., $M_{\text{eco}} = \sum_{q \in \mathcal{P}} m_q$, where $m_q$ denotes the maintenance score of package $q$. To incorporate dependency effects, we assume that the maintenance state of a package is not only determined locally, but also by the maintenance of the packages occurring in its dependency structure. To account for this, we replace the local maintenance state with a dependency-propagated maintenance measure. For a package $q$, we therefore express its dependency-propagated maintenance as $m_q^{\mathrm{eff}} = \sum_{p \in \mathcal{P}} o_{p,q} \cdot m_p$, where $m_p$ denotes the maintenance score of package $p$ and $o_{p,q} \in \{0,1\}$ indicates whether package $p$ appears in the transitive dependency closure of $q$, including $q$ itself as the root package. Substituting $m_q^{\mathrm{eff}}$ into the ecosystem state yields $M_{\text{eco}}^{\mathrm{eff}} = \sum_{q \in \mathcal{P}} \left( \sum_{p \in \mathcal{P}} o_{p,q} \cdot m_p \right)$. Reordering the finite sums leads to $M_{\text{eco}}^{\mathrm{eff}} = \sum_{p \in \mathcal{P}} \left( \sum_{q \in \mathcal{P}} o_{p,q} \right) \cdot m_p$. This reformulation shifts the perspective from aggregating contributions at the level of affected packages $q$ to expressing the ecosystem state in terms of contributions of individual packages $p$. The inner sum $\sum_{q \in \mathcal{P}} o_{p,q}$ captures how broadly package $p$ propagates through the ecosystem, i.e., the number of packages that directly or transitively depend on it, while the outer sum aggregates these contributions across all packages.
This representation makes it possible to isolate the contribution of a single package and analyze how changes to its maintenance score affect the overall ecosystem state. We define a scenario $\sigma$ as a specification of maintenance changes $\{\Delta m_p^{(\sigma)}\}_{p \in \mathcal{P}}$, where for each package $p \in \mathcal{P}$ the value $\Delta m_p^{(\sigma)}$ denotes the change in its maintenance score, e.g., by setting it to its maximum or minimum value in improvement and regression scenarios. The simulated maintenance change is given by $\Delta m_p^{(\sigma)} = m_p^{\text{after}} - m_p^{\text{before}}$, such that positive values indicate improvements in maintenance, while negative values indicate regressions. The resulting ecosystem impact $E_\sigma(p)$ of package $p$ is then given by:
\begin{align}
E_\sigma(p) = \Delta m_p^{(\sigma)} \cdot \sum_{q \in \mathcal{P}} o_{p,q}
\end{align}
which captures how a change in the maintenance score of $p$ propagates through the dependency network and affects other packages in the ecosystem. The resulting value $E_\sigma(p)$ can be positive or negative, where positive values correspond to improvements in ecosystem maintenance and negative values to regressions.
To obtain a comparable relative measure within a given scenario, we normalize this value by the total ecosystem impact induced by the same scenario:
\begin{align}
\hat{E}_\sigma(p) = \frac{E_\sigma(p)}{\sum_{r \in \mathcal{P}} E_\sigma(r)}
\end{align}
The resulting normalized ecosystem impact $\hat{E}_\sigma(p)$ represents the share of total ecosystem impact attributable to package $p$ under scenario $\sigma$. For scenarios with nonzero total induced impact, i.e., $\sum_{r \in \mathcal{P}} E_\sigma(r) \neq 0$, the normalization satisfies $\sum_{p \in \mathcal{P}} \hat{E}_\sigma(p) = 1$.
Aggregating these values over a selected subset of packages $S \subseteq \mathcal{P}$ yields the cumulative ecosystem impact:
\begin{align}
\sum_{p \in S} \hat{E}_\sigma(p) \ge \tau
\end{align}
where $\tau$ denotes a predefined ecosystem impact threshold (e.g., $\tau = 0.80$). This formulation allows us to prioritize packages based on their dependency-propagated ecosystem impact and to analyze how impact-driven support on a small subset of packages translates into ecosystem-level effects. This definition of ecosystem impact and the associated threshold-based selection criterion are used across all analyses~(RQ1, RQ4). Figure~\ref{fig:framework_overview} illustrates the overall analysis framework and how ecosystem impact modeling connects package selection strategies in RQ1 and RQ4 with the evaluation dimensions of RQ2 and RQ3.


\subsubsection{Impact-driven Support Strategy (RQ1)}
\label{sec:rq1-general-analysis}

To address RQ1, we analyze how a fixed ecosystem impact threshold of 80\% can be achieved by concentrating support on a small number of packages, following a Pareto-optimal allocation of ecosystem impact. For each package, ecosystem impact is derived from simulated maintenance changes to the \emph{Maintained} score on a bounded scale from $0$ to $10$. We define \emph{improvement} and \emph{regression} scenarios by setting the maintenance score of a package to the maximum value of $10$ and the minimum value of $0$. These endpoints are controlled bounds for comparing package selections, not predictions of realistic support outcomes.
For each package, we compute the normalized ecosystem impact $\hat{E}_\sigma(p)$ under both improvement and regression scenarios and rank packages in descending order for each case. For a given ecosystem impact threshold, e.g., $\tau=0.80$, we iteratively select top-ranked packages from each ranking until the cumulative normalized ecosystem impact satisfies Equation~(3) in each scenario.
The resulting package set is the union of the improvement-based and regression-based selections. We analyze the size and concentration of these package sets to determine how much of the total modeled ecosystem impact can be attributed to a small number of highly influential packages.
To assess whether the selection achieves ecosystem impact beyond chance, we compare the resulting package set against a size-matched random baseline. Let $S$ denote the set of packages selected under RQ1 with size $|S|$. We draw $N = 10{,}000$ equally sized sets randomly from the ecosystem using a fixed random seed to support reproducibility, and compute their aggregated ecosystem impact. From these Monte Carlo simulations, we estimate (i) the empirical probability that a random set achieves equal or greater impact than $S$, and (ii) a standardized effect size (z-score) relative to the simulated distribution.

\subsubsection{Maintainer Reach Analysis (RQ2)}
\label{sec:rq2-maintainer-reach}

For each selected package set, we analyze maintainer reach using repository ownership metadata. Individual repositories are counted as a single maintainer, whereas for organization-owned repositories, associated members are counted. Consequently, support strategies (RQ1, RQ4) can be compared by the total number of associated individuals and distinct maintainers reached. In addition, we report the number of single-maintainer packages, as these may indicate heightened concentration of responsibility and potential sustainability~risks.

\subsubsection{Metadata Accessibility and Exclusion Analysis (RQ3)}
\label{sec:rq3-metadata-coverage}

To address RQ3, we assess to what extent selected package sets are reachable and supportable through PyPI and repository metadata, based on the presence of contact information and donation platform links. In addition, we analyze systematic exclusion caused by missing repository links. For the impact-driven strategy (RQ1), we identify packages with comparable dependency-propagation reach, measured as $\sum_{q \in P} o_{p,q}$, that lack repository links and therefore cannot be assessed for maintenance activities. For existing mechanisms (RQ4), we examine the corresponding package sets and record entries without repository URL. Such cases indicate structural limitations and potential bias in metadata-driven support strategies.

\subsubsection{Comparison with Existing Support Mechanisms (RQ4)}
\label{sec:rq4-comparison}

For RQ4, we compare the package set selected under RQ1 with packages associated with existing support mechanisms. Specifically, we consider (i) PyPI packages covered by Tidelift~\cite{tidelift_packages_with_income}, (ii) PyPI packages supported by Ecosyste.ms~\cite{ecosystms_python_funds}, and (iii) PyPI packages linking to GitHub repositories whose owners have received at least one donation via GitHub Sponsors~\cite{github_sponsors}. This low threshold for considering a package supported by GitHub Sponsors is intentionally chosen to illustrate the results more clearly in Section~\ref{sec:results}. Tidelift's support model distinguishes between packages that are fully supported (referred to as \emph{lifted}) and packages that receive partial financial support. To provide a realistic estimate of Tidelift's ecosystem impact, our analysis focuses exclusively on \emph{lifted} packages.
For each support mechanism in RQ4, ecosystem impact is evaluated under the same improvement and regression scenarios used in RQ1, i.e., by simulating maximal maintenance improvement (score set to $10$) and maximal regression (score set to $0$) for all packages in the respective set. This ensures that differences in ecosystem impact reflect differences in package selection rather than differences in simulated maintenance changes. Finally, we apply the same analytical dimensions as in RQ1--RQ3.
\section{Results}
\label{sec:results}
Data collection started on February 17, 2026 and took approximately five days due to API rate limits imposed by GitHub and pepy.tech. The initial PyPI snapshot comprised 742,762 libraries retrieved via the PyPI simple index endpoint~\cite{pypi-simple}. During preprocessing, libraries that appeared only as dependency references but were not part of the initial ecosystem list were removed. This filtering step removed 1.5\% of libraries from a node perspective and 2.5\% from a dependency perspective. An additional 3.2\% of libraries were excluded due to errors when querying the PyPI JSON API~\cite{pypi_json_api}, mostly HTTP 404 responses for packages that could not be resolved via the detailed endpoint. The final dataset contains 718,750 libraries connected through 2,070,814 dependencies from 93,411 unique sources. This dataset forms the basis for all analysis conducted in this study. To ensure temporal alignment with the main dataset, package sets associated with Tidelift and Ecosyste.ms were collected separately on February 19, 2026 using a dedicated script that completed in under one minute.

\subsection{RQ1+RQ4: Comparative Ecosystem Impact Analysis}

This subsection presents the ecosystem impact results for our impact-driven strategy (RQ1) and existing support mechanisms (RQ4).
Table~\ref{tab:ecosystem_support_comparison} summarizes the normalized, dependency-propagated ecosystem impact achieved by each strategy under simulated improvement and regression scenarios. For each strategy, we report the resulting impact share, the number of selected packages and their proportion relative to the full PyPI ecosystem.

\begin{table*}[ht]
\caption{Comparison of ecosystem support strategies under improvement and regression scenarios}
\centering
\footnotesize
\setlength{\tabcolsep}{6pt}
\renewcommand{\arraystretch}{1.15}
\resizebox{\textwidth}{!}{%
\begin{tabular}{l|c|cc|cc}
\hline
\textbf{Strategy} &
\makecell{\textbf{Research} \\ \textbf{question}} &
\makecell{\textbf{Improvement} \\ \textbf{impact}} &
\makecell{\textbf{Regression} \\ \textbf{impact}} &
\makecell{\textbf{Number of} \\ \textbf{packages}} &
\makecell{\textbf{\% of} \\ \textbf{ecosystem}} \\
\hline
Impact-driven     & RQ1 & 80.01\% & 82.15\% & 730    & 0.10\% \\
Tidelift (lifted) & RQ4 & 22.66\% & 38.65\% & 277    & 0.04\% \\
Ecosyste.ms       & RQ4 & 25.90\% & 38.00\% & 97     & 0.01\% \\
GitHub Sponsors   & RQ4 & 27.89\% & 36.71\% & 20,433 & 2.84\% \\
\hline
\end{tabular}
}
\label{tab:ecosystem_support_comparison}
\end{table*}

The impact-driven strategy achieves 80.01\% improvement impact and 82.15\% regression impact by selecting 730 packages, corresponding to 0.10\% of the ecosystem. This confirms that ecosystem impact in PyPI is highly concentrated: a very small subset of packages accounts for a large share of dependency-propagated ecosystem impact. To assess whether this result could arise under uniform allocation, we compare the selected set against 10{,}000 size-matched random selections. The impact-driven set exceeds the mean random impact by 161.04 standardized units in the improvement scenario and 295.51 in the regression scenario, with no random set achieving comparable impact ($p < 0.0001$).

For the existing support mechanisms (Tidelift, Ecosyste.ms, and GitHub Sponsors), ecosystem impact is computed under the same improvement and regression scenarios as in RQ1, ensuring that differences reflect package selection rather than differences in simulated maintenance changes.
For Tidelift, we consider only lifted packages, which receive full support. The lifted set comprises 277 packages (0.04\% of the ecosystem) and achieves 22.66\% improvement impact and 38.65\% regression impact. Despite including a substantial number of packages, its ecosystem impact remains lower than for the impact-driven strategy, which is consistent with selection criteria that do not optimize dependency-propagated impact.
Ecosyste.ms covers 97 packages (0.01\% of the ecosystem) and achieves 25.90\% improvement and 38.00\% regression impact. Despite its relatively small size, its ecosystem impact is comparable to Tidelift, suggesting partial alignment with impactful packages.
GitHub Sponsors, defined here as PyPI packages whose linked GitHub repository owner has received at least one sponsorship, includes 20{,}433 packages (2.84\% of the ecosystem), making it the largest set among the analyzed mechanisms. Despite this size, its ecosystem impact remains lower than for the impact-driven strategy (27.89\% improvement, 36.71\% regression), which is consistent with sponsorship decisions being shaped by individual preferences for particular developers or projects rather than dependency-propagated ecosystem impact.

\subsection{RQ2+RQ4: Comparative Maintainer Reach Analysis}

This subsection analyzes how the package sets identified in RQ1 and RQ4 translate into maintainer reach. To contextualize maintainer-related metrics, Table~\ref{tab:rq2_maintainer_reach} combines ecosystem impact results with maintainer metrics. For each strategy, we report (i) the number of packages, (ii) ecosystem impact shares under improvement and regression scenarios, (iii) the total number of associated repository individuals, (iv) the number of distinct maintainers, and (v) the proportion of single-maintainer packages. Total individuals count all repository members linked to the selected packages of a given strategy, while distinct maintainers represent unique individuals across all selected packages within that strategy. Single-maintainer packages indicate cases where maintenance responsibility is concentrated on a single individual.

\begin{table*}[ht]
\caption{Maintainer reach across ecosystem support strategies}
\centering
\small
\setlength{\tabcolsep}{6pt}
\renewcommand{\arraystretch}{1.15}
\resizebox{\textwidth}{!}{
\begin{tabular}{l|c|c|cc|cc|c}
\hline
\textbf{Strategy} &
\makecell{\textbf{Research} \\ \textbf{question}} &
\makecell{\textbf{Number of} \\ \textbf{packages}} &
\makecell{\textbf{Improvement} \\ \textbf{impact share}} &
\makecell{\textbf{Regression} \\ \textbf{impact share}} &
\makecell{\textbf{Total} \\ \textbf{individuals}} &
\makecell{\textbf{Distinct} \\ \textbf{maintainers}} &
\makecell{\textbf{Single-maintainer} \\ \textbf{packages}} \\
\hline
Impact-driven     & RQ1 & 730    & 80.01\%  & 82.15\%     & 31,930    & 12,255 & 311 (42.6\%)   \\
Tidelift (lifted) & RQ4 & 277    & 22.66\%  & 38.65\%     & 4,070     & 1,119  & 101 (36.5\%)   \\
Ecosyste.ms       & RQ4 & 97     & 25.90\%  & 38.00\%     & 1,880     & 1,001  & 23 (23.7\%)    \\
GitHub Sponsors   & RQ4 & 20,433 & 27.89\%  & 36.71\%     & 3,204,168 & 6,504  & 5,361 (26.2\%) \\
\hline
\end{tabular}
}
\label{tab:rq2_maintainer_reach}
\end{table*}


Table~\ref{tab:rq2_maintainer_reach} shows that maintainer reach differs substantially across strategies and does not scale proportionally with ecosystem impact. The impact-driven strategy (RQ1) achieves 80.01\% improvement impact and 82.15\% regression impact with 730 packages, reaching 31,930 individuals and 12,255 distinct maintainers. While this reflects a broad maintainer reach, it is primarily driven by the larger number of selected packages rather than an inherent property of the prioritization approach.
For Tidelift (lifted), 277 packages result in 22.66\% improvement impact and 38.65\% regression impact, reaching 4,070 individuals across 1,119 distinct maintainers. Compared to the impact-driven strategy, maintainer reach remains substantially lower, reflecting the smaller size of the selected package set.
Ecosyste.ms covers 97 packages and achieves 25.90\% improvement impact and 38.00\% regression impact, reaching 1,880 individuals across 1,001 distinct maintainers. Despite its smaller package set compared to Tidelift, its distinct maintainer reach is relatively close, suggesting a higher density of maintainers per package, but without a corresponding increase in ecosystem impact.
GitHub Sponsors includes 20,433 packages, corresponding to 27.89\% improvement impact and 36.71\% regression impact. The set is associated with 3,204,168 individuals but only 6,504 distinct maintainers, indicating substantial overlap in repository membership and the presence of large organizational repositories. Despite this broad maintainer reach, the associated ecosystem impact remains comparatively limited.
Across strategies, the proportion of single-maintainer packages ranges from 23.7\% (Ecosyste.ms) to 42.6\% (impact-driven), with 26.2\% for GitHub Sponsors and 36.5\% for Tidelift (lifted). As single-maintainer projects are often more vulnerable to sustainability risks, higher proportions indicate packages that may particularly benefit from additional support.

\subsection{RQ3+RQ4: Comparative Metadata Accessibility Analysis}

This subsection examines the feasibility of operationalizing ecosystem support strategies based on the availability of actionable metadata. Table~\ref{tab:rq3_metadata_accessibility} reports, for each strategy, the number of selected packages, the proportion providing contact information, the proportion linking to donation platforms, and their overlap, as well as the number of packages excluded due to missing repository links. Repository links are required to assess maintenance activities and apply the dependency-propagated impact model. For the modeled strategy (RQ1), exclusions occur when packages with similar dependency-propagation reach as the selected sets do not provide a repository URL. For externally defined mechanisms (RQ4), they reflect packages in the supported sets without resolvable repository links.

\begin{table*}[ht]
\caption{Metadata accessibility and exclusion across ecosystem support strategies}
\centering
\small
\setlength{\tabcolsep}{6pt}
\renewcommand{\arraystretch}{1.15}
\resizebox{\textwidth}{!}{
\begin{tabular}{l|c|c|ccc|c}
\hline
\textbf{Strategy} &
\makecell{\textbf{Research} \\ \textbf{question}} &
\makecell{\textbf{Number of} \\ \textbf{packages}} &
\makecell{\textbf{Packages with} \\ \textbf{contact info (1)}} &
\makecell{\textbf{Packages with} \\ \textbf{donation link (2)}} &
\makecell{\textbf{Packages with} \\ \textbf{(1) and (2)}} &
\makecell{\textbf{Excluded packages} \\ \textbf{(no repository link)}} \\
\hline
Impact-driven     & RQ1 & 730    & 705 (96.6\%)    & 245 (33.6\%)     & 241 (33.0\%)    & 26 (1.8\%)  \\
Tidelift (lifted) & RQ4 & 277    & 258 (93.1\%)    & 213 (76.9\%)     & 213 (76.9\%)    & 17 (6.1\%)  \\
Ecosyste.ms       & RQ4 & 97     & 89 (91.8\%)     & 57 (58.8\%)      & 57 (58.8\%)     &  7 (7.2\%)  \\
GitHub Sponsors   & RQ4 & 20,433 & 19,650 (96.2\%) & 20,431 (99.99\%) & 19,648 (96.2\%) &  0 (0.0\%)  \\
\hline
\end{tabular}
}
\label{tab:rq3_metadata_accessibility}
\end{table*}


As shown in Table~\ref{tab:rq3_metadata_accessibility}, the impact-driven strategy (RQ1) exhibits high contact information coverage (96.6\%), indicating that maintainers of the selected packages are generally reachable. In contrast, donation link availability is substantially lower (33.6\%), with 33.0\% of packages providing both contact information and donation links. A total of 26 packages (1.8\%) are excluded due to missing repository links, indicating limited but non-negligible constraints for applying the dependency-propagated impact model.
For externally defined mechanisms~(RQ4), metadata accessibility varies across support sets. Tidelift (lifted) shows similarly high contact coverage (93.1\%) but substantially higher donation link availability (76.9\%), with all packages providing donation links also providing contact information. However, 17 packages (6.1\%) lack repository URLs and are therefore excluded.
Ecosyste.ms exhibits slightly lower contact coverage (91.8\%) and moderate donation link availability (58.8\%), with identical overlap between contact information and donation links. A total of 7 packages (7.2\%) are excluded due to missing repository links.
GitHub Sponsors shows near-complete donation-link coverage (99.99\%) by construction, as inclusion in the set requires at least one sponsorship on the platform. Contact information is available for 96.2\% of packages, with the same proportion providing both. No packages are excluded, as all entries link to resolvable repositories.
\section{Discussion}

\subsection{Structural Concentration and Support Efficiency}

Our findings show that ecosystem impact in the PyPI dependency network is highly concentrated. Under the dependency-propagated model, a large share of the overall impact (approximately 80\%) can be attributed to a small subset of packages (0.10\% of the ecosystem) when prioritized based on impact. This confirms that maintenance-related influence propagates unevenly through the dependency network, with relatively few packages affecting large portions of the ecosystem.
These results align with prior observations of concentration effects in package dependency networks, where a small number of packages accumulate substantial influence through a disproportionate number of dependents~\cite{decan2019empirical}. However, our findings extend this work by quantifying how structural concentration translates into dependency-propagated maintenance impact under simulated interventions.
From a support perspective, this concentration implies that targeting a small fraction of packages can achieve substantial ecosystem-level effects. At the same time, it does not necessarily reduce maintainer reach, as even small sets of high-impact packages can involve large maintainer populations.

\subsection{Structural Alignment of Existing Mechanisms}

The comparison with existing support mechanisms reveals notable variation in their alignment with dependency-propagated ecosystem impact. Ecosyste.ms shows comparatively strong alignment with high-impact packages despite its small size, whereas Tidelift’s lifted set and GitHub Sponsors achieve lower impact shares despite larger package sets. This suggests that selection criteria driven by subscription participation or sponsorship preferences do not necessarily align with impact-driven prioritization.
These observations do not imply that existing mechanisms are ineffective, but rather that different allocation logics optimize for different objectives. While impact-driven selection prioritizes packages based on their ecosystem-wide effects under maintenance changes, existing mechanisms reflect factors such as user demand, preferences, or visibility. This is consistent with prior findings that funding and sponsorship in OSS ecosystems are shaped by project-level characteristics and social signals, which do not necessarily align with ecosystem-level impact considerations~\cite{overney2020donations,shimada2022sponsors}.

\subsection{Maintainer Reach and Spillover Effects}

Maintainer reach analysis (RQ2) shows that structural impact and social footprint are related but distinct dimensions. Strategies with similar ecosystem impact can reach very different numbers of maintainers, and large maintainer counts may reflect organizational repositories rather than increases in distinct individuals. This distinction between structural influence and social footprint complements prior work emphasizing the role of maintainer capacity and human infrastructure in OSS sustainability~\cite{linaaker2024sustaining}. The prevalence of single-maintainer packages also indicates that impactful or externally supported packages rely on concentrated maintenance responsibility, which prior work links to elevated sustainability risk~\cite{avelino2016novel} and may therefore benefit disproportionately from additional support. Moreover, maintainers of high-impact packages often contribute to multiple libraries, suggesting that supporting them may indirectly benefit additional packages beyond the selected set. Quantitative evidence for these developer-level spillover effects is provided in the replication package.

\subsection{Metadata Accessibility as an Operational Constraint}

The metadata accessibility analysis (RQ3) shows that contact information is widely available across all package sets, including impact-driven strategies (RQ1) and external support mechanisms (RQ4). In practice, this indicates that maintainers of relevant packages are generally reachable, enabling direct communication and coordination of support efforts. In contrast, donation-link availability varies more strongly: externally defined mechanisms exhibit high financial metadata coverage, while packages selected under RQ1 show more heterogeneous donation-link availability. This implies that, despite being reachable, many maintainers cannot be easily supported through established financial channels, limiting the immediate feasibility of monetary interventions. Missing repository links represent an additional operational constraint by limiting maintenance activity assessment and impact simulation. The primary bottlenecks therefore lie in the availability of financial support channels and repository links. Metadata accessibility thus represents an independent feasibility dimension for ecosystem support strategies. These findings align with prior studies on metadata accessibility, which report high availability of contact information but limited adoption of donation platforms and repository links across PyPI libraries~\cite{tsakpinis2024analyzing,tsakpinis2025analyzing,tsakpinis2026analyzing}.

\subsection{Comparison of Ecosystem Impact Metric to PageRank Score}

To assess the added value of our dependency-propagated ecosystem impact metric, we compare it against PageRank as an established measure of structural importance in dependency networks. While PageRank provides a global ranking of libraries based on their position in the dependency graph, our metric ranks libraries according to the extent to which changes in their maintenance state propagate and influence the global ecosystem state.

\textbf{Divergence in prioritized packages.}
We first analyze the overlap between the package set obtained in RQ1 with the \text{top-$k$} packages selected by PageRank ($k = 730$). The Jaccard similarity between the two sets is 41.06\%, indicating that less than half of the selected packages are shared. The symmetric difference further reveals that 305 packages are selected exclusively by each method. This divergence shows that the two approaches prioritize fundamentally different subsets. In particular, packages selected by our metric but not by PageRank reflect cases where strong dependency-propagated impact is not captured by structural centrality, while PageRank-specific selections correspond to structurally central packages whose ecosystem impact is comparatively lower. These findings indicate that PageRank alone is insufficient to capture the effects of maintenance changes on the ecosystem.

\textbf{Performance under fixed support budgets.}
To compare both approaches under identical resource constraints, we compare the package set obtained in RQ1 with the \text{top-$k$} packages selected by PageRank ($k = 730$) and evaluate the resulting total ecosystem impact. The impact-driven strategy achieves a higher cumulative improvement impact (80.01\%) than the PageRank-based selection (69.71\%), indicating that it is more effective in identifying packages whose maintenance improvements yield larger ecosystem-wide benefits. For regression scenarios, the PageRank-based selection achieves slightly higher total impact (84.64\% vs.\ 82.15\%), suggesting that structural centrality is a strong proxy when modeling worst-case degradation effects. This difference highlights that PageRank effectively identifies structurally central packages whose maintenance degradation has large ecosystem-wide effects, while our proposed metric more effectively identifies intervention points for improving ecosystem maintenance. Overall, these results show that the choice of prioritization strategy influences the effectiveness of support allocation depending on whether the objective is mitigation of risk or maximization of improvement.

\textbf{Correlation between PageRank and ecosystem impact.}
To quantify the relationship between structural centrality and dependency-propagated ecosystem impact, we compute Spearman rank correlations between PageRank scores and the normalized ecosystem impact values $\hat{E}_{\sigma}(p)$ under improvement and regression scenarios. Spearman correlation is used as the primary metric, as PageRank and our impact metric are ultimately used for prioritizing libraries, making agreement in ranking the most relevant notion of similarity~\cite{de2016comparing}. The correlation is low for improvement impact ($r_s = 0.17$) and slightly higher for regression impact ($r_s = 0.28$), indicating only weak agreement between the rankings. This suggests that PageRank is not a reliable proxy for ecosystem impact, particularly when considering maintenance improvements. The higher correlation in regression scenarios is consistent with the observation that maintenance degradation in structurally central packages propagates widely through the dependency network.

We also compute Pearson correlations to assess linear relationships between PageRank scores and the normalized ecosystem impact values $\hat{E}_{\sigma}(p)$ under improvement and regression scenarios. Here, we observe a moderate correlation for improvement impact ($r_p = 0.34$) and a strong correlation for regression impact ($r_p = 0.87$). This indicates that PageRank aligns more closely with regression impact in terms of magnitude, reflecting that structurally central packages tend to dominate worst-case degradation scenarios. However, Pearson correlation captures linear association rather than agreement in prioritization and is sensitive to skewed distributions and outliers~\cite{de2016comparing}, which are characteristic of dependency networks~\cite{decan2019empirical}. Therefore, while Pearson correlation helps explain the strong performance of PageRank in regression scenarios observed previously, it should not be overemphasized when evaluating selection strategies, where rank-based agreement is more relevant.

\textbf{Implications for resource allocation.}
A key distinction between PageRank and our proposed metric lies in how they support decision-making. PageRank provides a static ranking of libraries and can be used as a heuristic for selecting influential packages. However, it does not account for how different levels of maintenance intervention affect the ecosystem. In contrast, our impact metric explicitly models the effect of changes in maintenance scores~($\Delta m$), enabling reasoning about the consequences of different intervention magnitudes. This makes it possible to frame support allocation as an optimization problem, where decisions are not limited to which libraries to support, but also how to distribute resources across them. While the present study focuses on threshold-based selection, the underlying formulation supports more general allocation strategies that trade off the number of supported packages against the magnitude of maintenance improvements and regressions.

\textbf{Summary.}
These results show that PageRank and our ecosystem impact metric capture related but distinct aspects of ecosystem behavior. While PageRank is effective in identifying structurally central packages and modeling worst-case degradation scenarios, it fails to account for maintenance dynamics and intervention effects. The proposed metric complements structural centrality by incorporating maintenance changes and their propagation, providing a more expressive basis for prioritizing and allocating support in software~ecosystems.

\subsection{Practical Implications}

Taken together, the results suggest that dependency-propagated ecosystem impact modeling provides a practical and transparent approach for prioritizing support under explicit structural criteria. Because ecosystem support is multi-dimensional, allocation decisions should jointly consider ecosystem impact, maintainer reach, responsibility concentration, and metadata accessibility. A key implication of the impact formulation is that it enables optimization-based resource allocation. Different configurations of supported packages can achieve the same ecosystem impact threshold while varying in the number of selected libraries and the maintenance effort required per package. This defines a trade-off space in which multiple configurations are feasible, with the appropriate choice depending on context-specific constraints such as available resources, desired maintainer coverage, or operational feasibility. The proposed framework is not intended to replace existing mechanisms, but to make such trade-offs explicit and measurable. Potential practitioners include ecosystem stewards, funding bodies, companies managing open source risk in their software supply chains, and platform providers operating support programs. For these actors, dependency-propagated impact modeling can serve as a decision-support layer that complements existing processes for evaluating and guiding open source support and funding interventions~\cite{osborne2024toolkit}.

\section{Threats to Validity}
We structure threats to validity into the following four aspects~\cite{runeson2009guidelines}:

\textbf{Construct validity:}
Ecosystem impact is operationalized as the normalized, dependency-propagated effect of simulated changes in the OpenSSF \textit{Maintained} score. Because this metric captures only selected maintenance aspects and may not fully reflect project health, multidimensional assessments using suitable CHAOSS metrics could yield different rankings~\cite{linaaker2026assessing}. The model also excludes external signals such as industry adoption and usage beyond the ecosystem. Maintainer reach and metadata accessibility are inferred from repository ownership, member metadata, contact information, and donation links, which may not fully represent active contributors or the practical feasibility of direct contact and financial support. However, prior work suggests that e-mail availability is a reasonable reachability proxy, supported by validated deliverability, an undeliverable rate of approximately 11\%, and the prevalence of established e-mail domains~\cite{tsakpinis2026authors,tsakpinis2026analyzing}.

\textbf{Internal validity:} 
Our analysis is based on a cross-sectional snapshot of PyPI and GitHub metadata. Because dependency structures and maintenance signals evolve over time, absolute impact values may change. Also, the linear model and 0/10 bounds support controlled comparison, but omit richer dependency effects and do not establish the effects of actual support interventions.

\textbf{External validity:} 
The study focuses on the PyPI ecosystem and publicly accessible metadata, limiting generalizability to ecosystems with different dependency structures, package management practices, or metadata availability (e.g., npm, Maven). In addition, repository-level analysis is restricted to GitHub, as PyPI libraries predominantly link to GitHub rather than alternatives such as GitLab or Bitbucket~\cite{tsakpinis2024analyzing}. The methodology could nevertheless be applied to other ecosystems if comparable metadata is available.

\textbf{Reliability:} 
Details on the replication package are provided in Section~\ref{sec:data_availabaility}. Since ecosystem and repository metadata evolve over time, re-collected data may yield different results.

\section{Conclusion and Future Work}

This paper introduced a dependency-aware ecosystem impact model that quantifies how simulated maintenance changes propagate through the PyPI dependency network. By combining dependency propagation with OpenSSF Scorecard maintenance signals, we operationalized ecosystem impact as a normalized dependency-propagated metric and framed ecosystem support allocation as a constrained selection problem. Our results show that ecosystem impact in PyPI is highly concentrated: a large share of the modeled impact can be attributed to a small fraction of packages when prioritized based on dependency-propagated impact. Benchmarking existing mechanisms shows substantial variation in their alignment with ecosystem impact. Maintainer reach and metadata accessibility further emerge as complementary dimensions shaping the practical feasibility of ecosystem support strategies.

Future work should extend the framework longitudinally to monitor evolving dependency structures and maintenance signals, enabling continuous ecosystem monitoring and earlier detection of emerging risks. Refining the propagation model to identify low-maintenance libraries near root packages as potential structural bottlenecks may improve intervention effectiveness by targeting points where degraded maintenance could limit upstream benefits. Additionally, the proposed impact formulation supports optimization-based support allocation, where resources can be distributed across libraries and intervention magnitudes to maximize ecosystem-level outcomes under practical constraints. Finally, future work should test multidimensional maintenance indicators and richer propagation models, validate modeled effects against observed support outcomes, and study other ecosystems.
\section{Data Availability}
\label{sec:data_availabaility}
All data and analysis artifacts required to reproduce the results of this study are publicly available on Figshare under a CC-BY 4.0 license~\cite{tsakpinis_pretschner_2026}. The package further includes the data collection pipeline, allowing re-execution of the study with updated data.

\bibliography{technical_paper}

\end{document}